# Detecting Syllable-Level Pronunciation Stress with A Self-Attention Model[*]

WANG Weiying[1], NAKAJIMA Akinori[2]

[1]School of Environment and Society, Tokyo Institute of Technology, Tokyo 152-8550, Japan.
wang.w.al@m.titech.ac.jp

[2]VoicePing Inc., Tokyo 105-7510, Japan. info@nakajima-akinori.com

**Abstract:** One precondition of effective oral communication is that words should be pronounced clearly, especially for non-native speakers. Word stress is the key to clear and correct English, and misplacement of syllable stress may lead to misunderstandings. Thus, knowing the stress level is important for English speakers and learners. This paper presents a self-attention model to identify the stress level for each syllable of spoken English. Various prosodic and categorical features, including the pitch level, intensity, duration and type of the syllable and its nuclei (the vowel of the syllable), are explored. These features are input to the self-attention model, and syllable-level stresses are predicted. The simplest model yields an accuracy of over 88% and 93% on different datasets, while more advanced models provide higher accuracy. Our study suggests that the self-attention model can be promising in stress-level detection. These models could be applied to various scenarios, such as online meetings and English learning.

## 1 Introduction

Effective oral communication is essential for daily and formal conversations. However, spoken messages can be misunderstood if words are not pronounced distinctly. For English pronunciation, syllable stress (also called word stress or lexical stress) is an important element because English makes more use of stress than many other languages (Ferrer et al., 2015). Formally speaking, English is a stress-timed language, which means that stressed syllables are longer and have a higher pitch, while unstressed syllables are shorter and have a lower pitch. Here are some word examples, where the primarily stressed syllables are capitalized and bolded:

<div style="text-align:center">

over-**COME**
e-**MO**-tion
**UN**-der-wear

</div>

Generally, for words with two or more syllables, there should be only one primarily stressed syllable, and there can be one or multiple secondarily stressed or unstressed syllables. Native English speakers identify words not only by the pronunciation of sounds but also by the stress patterns. Misplaced syllabic stress may change the meaning of the word, and eventually causes misunderstanding. Thus, awareness of correct syllable stress is important for English speakers.

Many features have been used to detect syllable stress automatically. Among the state-of-the-art models, three features were most commonly used: the pitch level (f0 frequency), energy, and duration. Depending on the model, these three measurements were either evaluated over the syllable or the vowel of the syllable. The vowel of a syllable is also called the nucleus of the syllable, and one syllable has exactly one vowel (nucleus). Many previous models treated the syllable stress detection as a binary classification task: a syllable is either stressed or unstressed. Some models only predicted one stressed syllable for a spoken word. However, this labeling assumption is inappropriate since non-native speakers may mistakenly stress several syllables in a word (Ferrer et al., 2015). A comprehensive prediction system should make predictions for each word syllable and include three types of stress levels: unstressed, primarily stressed, and secondarily stressed. This setting makes the detection task challenging.

---

[*] This work was conducted while the first author was doing the internship at VoicePing Inc. Source codes are available at https://github.com/wangweiying303/stress-detection-model.

In this article, we apply a self-attention model to predict syllable stress as one of the three levels. To address the limitation of datasets, speech audios are automatically generated and segmented with deep learning models and annotated with a dictionary, assuming that all syllables are correctly stressed. This paper has five sections. In the next section (2), the general research flow and the self-attention model are described. In Section 3, the datasets, preprocessing, and feature extraction are introduced. Section 4 presents the experiments and findings, and Section 5 concludes the study.

# 2 Research Flow and the Self-Attention Model

For English, different studies have used different datasets for stress detection. For example, the ISLE corpus is a popular one that contains English audio from 46 non-native speakers (23 German and 23 Italian) with 7834 utterances. Other researchers used their own datasets. However, these annotated datasets are not available to us. Given the lack of annotated datasets, we synthesized a large amount of audio data for training and testing the model. This section presents the research flow and the model specification.

## 2.1 Research Flow

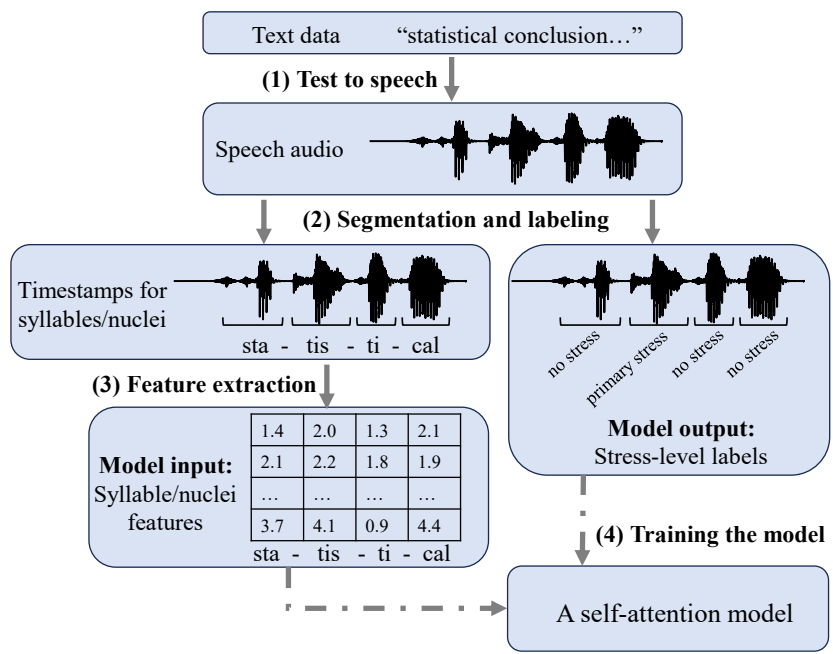

Fig. 1 Flowchart of the study.

The research flowchart is shown in Fig. 1. We first convert a large number of texts into humanlike speech audios using a deep-learning model (step (1) in Fig. 1), assuming that word syllables are correctly stressed the same as the stress level from a dictionary. Second, for any speech audio, timestamps are tagged for syllables/nuclei, and each syllable is labeled a stress level using the dictionary (step (2) in Fig. 1). Third, syllable features are extracted from the audio data (step (3) in Fig. 1). Finally, a self-attention model is applied to predict the stress-level of each syllable using the extracted features (step (4) in Fig. 1). Standard pronunciation audios are generated using the Azure Text-To-Speech API (TTS hereafter). Speech audios are segmented with the Azure Pronunciation Assessment API (PA hereafter).

## 2.2 Model Specification

A self-attention model is applied to predict the stress level. When predictions are made, self-attention models look into not only the features of the target syllable but also the features of syllables around the target. Many previous models consider more or less the features of the proceeding or following syllable when predicting the stress level of the target syllable (Tepperman and Narayanan, 2005; Ferrer et al., 2015; Ramanathi et al., 2019). The self-attention model is in nature a solution for this purpose. Moreover, although stressed syllables are generally longer and have higher pitches, the pronunciation pattern of a syllable also depends on the vowels and consonants it contains. Thus, the prosodic features of stressed syllables may vary depending on the vowel type. Self-attention models address this problem by embedding the syllable type into a vector and taking it as the input. The self-attention model applied in this study is in a similar framework to other transformer-based models (e.g., the vision transformer by Dosovitskiy et al., 2020). Fig. 2 shows the model architecture.

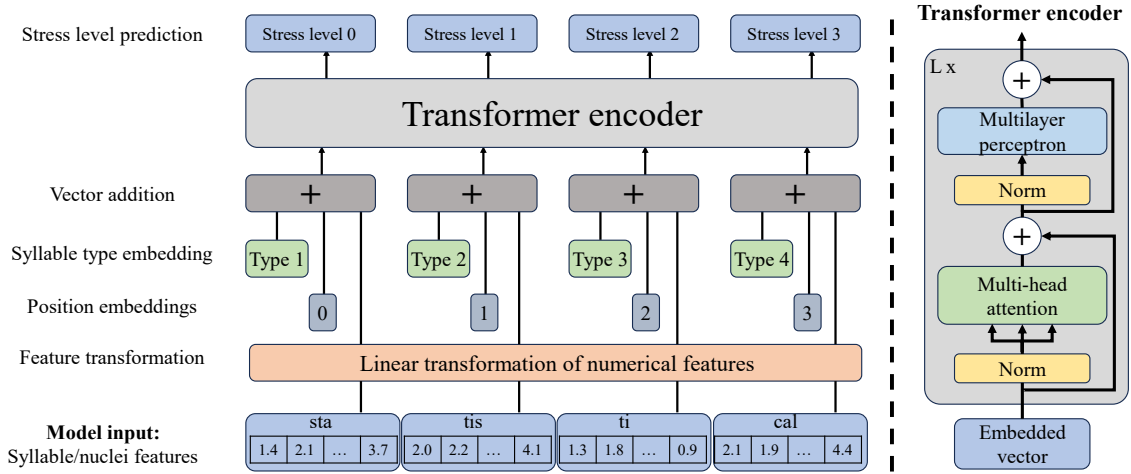

Fig. 2 Overview of the self-attention model for syllable stress detection.

The self-attention model consists of two components: feature embedding layers and a transformer encoder. For each syllable in a word, the feature embedding layers convert the features (including prosodic features, syllable type, and position of syllables in the word) into a vector with the length of $D$:

$$v_i = E_{pos}(p_i) + E_{type}(t_i) + x_i \boldsymbol{C} \tag{1}$$

where $p_i$ is the position of the $i$-th syllable in the word, $E_{pos}(p_i)$ is the position embedding of length $D$ for the syllable, $t_i$ is the type of the syllable, $E_{type}(t_i)$ is the type embedding of length $D$ for the syllable, $x_i \in \mathbb{R}^{1 \times K}$ is the feature vector of the syllable, and $\boldsymbol{C} \in \mathbb{R}^{K \times D}$ is the linear projection that covert the vector $x_i$ of length $K$ to a vector with the length of $D$. For each word, the syllable vectors are stacked into a matrix ($v$). To handle different numbers of syllables for different words, each word is supposed to have seventeen syllables (i.e., $v$ has the length of seventeen). For words with fewer syllables, the syllable matrix is padded with zero vectors. The syllable matrix $v$ is the standard input for the transformer encoder. It includes a self-attention mechanism, and the padded zero vectors are masked when the mechanism is applied. The transformer encoder outputs the probabilities of stress level for each syllable in a word. The details of the transformer encoder are out of the scope of this paper. Interested readers may read the works by Vaswani et al. (2017) for more details.

# 3 Data Collection and Feature Extraction

As mentioned in Subsection 2.1, we apply the Azure TTS and PA modules to generate and segment audio datasets. Syllable stress is labeled with the CMU Pronunciation Dictionary[1], a dictionary for North American English. A Python package (namely "parselmouth") is employed to extract audio features.

## 3.1 Data Collection with Azure Modules

Azure TTS is an API that converts text into humanlike synthesized speech. The "LibriSpeech ASR" corpus[2] is downloaded and texts from the corpus are used as the input to the Azure TTS. To extract syllable (and nuclei) features from any audio data, the timestamps for syllables (and nuclei) are necessary. In the old times, timestamps were annotated by professionals, which was tedious and time-consuming. In the last two decades, researchers have applied several types of forced alignment algorithms to segment audio into syllable slices (e.g., Hidden Markov Models). Conventional models are less accurate and manual corrections are needed. However, in real use cases, manual corrections are not possible (Yarra and Ghosh, 2022). In this research, we use the timestamp labeled by the Azure PA, which is powered by a deep learning model and presumably may provide better accuracies.

## 3.2 Feature Extraction and Normalization

With the timestamps prepared for syllables and nuclei, various features can be extracted. Since stressed syllables are longer and have higher pitch and energy, we focus on thirteen measurements characterizing these features. There are six numerical measurements over each syllable: the average pitch level; the maximal pitch level; the duration of the syllable segment with detectable pitch level; the average intensity level; the maximal intensity level; and the duration of the syllable segment. The same numerical measurements are extracted over the nucleus of the syllable. These numerical features can be easily obtained with the Python package named "parselmouth". In addition, considering that different types of syllables are stressed differently because of their nucleus types, the nucleus type is taken into account. As a result, the model inputs include twelve numerical features and one categorical feature. The Azure PA module returns the nucleus (the vowel phoneme) type of each syllable, and there are sixteen types of nuclei from the PA returns.

All numerical features are normalized so that the average of any feature is zero at the sentence level (i.e., average values are subtracted from the feature vectors). Normalization is performed such that audio features are independent of the speech rate, pitch level, etc., of the speaker.

---

[1] The CMU dictionary is available at http://www.speech.cs.cmu.edu/cgi-bin/cmudict.
[2] The dataset is available at https://www.openslr.org/12.

# 4 Experiments

We evaluate the learning ability of the self-attention model by varying feature size, the dataset for training and testing, and the model complexity. The benchmark models are the Ordinal Regression model (OR hereafter) and the Random Forest classifier (RF hereafter). They are frequently applied models in classification tasks. Some other classifiers, such as the decision tree model and the Support Vector Machine, do not show superior performance and are not presented here.

## 4.1 Setup

Three datasets are collected and applied in the study. The first dataset is the read English speech named "train-clean-100" audio from the "LibriSpeech ASR" corpus ("Read Data" hereafter) mentioned in Subsection 3.1. It contains 30,453 utterances from 247 readers with various English accents. The second and third datasets are synthesized with the Azure TTS module using the same texts from the corpus. The second data is spoken by a male virtual speaker ("en-US-ChristopherNeural" from Azure TTS) and the third one by a female speaker ("en-US-JennyNeural"). The second and the third datasets are combined into one data named "Generated Data" hereafter. When training and testing models, only words with two or more syllables are included. The Azure PA module may not return the syllable-level information every time. When there is a mismatch between the Azure PA return and the CMU pronunciation dictionary in terms of the number of syllables in a word, the audio data is excluded. For the Read (Generated) Data, we obtained 28,970 (56,318) sentences and 142,043 (283,971) non-unique words with multiple syllables. 70% of datasets are used for training and 30% of datasets are used for testing.

The complexity of the self-attention model is controlled by the vector size $D$, the number of heads, and layers of the transformer encoder. Two different sizes of models are trained and compared (Table 1).

Table 1 Details of the two models.

| Model | Vector Size $D$ | Heads | Layers |
|---|---|---|---|
| Model-medium | 5 | 6 | 3 |
| Model-large | 10 | 12 | 6 |

We adopt a progressive manner when inputting audio features into models. In the first step, models are trained and evaluated using the six numerical features over syllables ("Syllable Numerical" hereafter). Secondly, models are examined with all the twelve numerical features over syllables and nuclei ("Syllable/Nucleus Numerical" hereafter). Lastly, all the numerical features and the categorical features (i.e., the type of syllable) are input into the models ("All Features" hereafter). Since the stress levels for each syllable type are highly unbalanced, some syllables might be always predicted as stressed or unstressed when syllable type is embedded. To cope with this problem, stress level is weighted for each syllable type by:

$$w_c(s) = \left(\frac{p_c(s)}{\max(p_c(i))}\right)^{0.7} s \in (s_0, s_1, s_2), i = s_0, s_1, s_2 \qquad (2)$$

where $w_c(s)$ is the weight for stress level $s$ for nucleus type $c$; $p_c(s)$ is the proportion of syllables with stress level $s$ for nucleus type $c$; $\max(p_c(i))$ is the highest proportion of stress level of this nucleus type. $s_0, s_1, s_2$ are the non-stress, primary stress, and secondary stress respectively. Fig. 3 shows the proportion of stress levels for each type of syllable (nucleus types) before and after the weights are assigned. The stress levels seem to be more evenly distributed when weights in

Equation (2) are assigned. In the following, when syllable types are embedded as input, the loss is weighted by $w_c(s)$ for stress level for training.

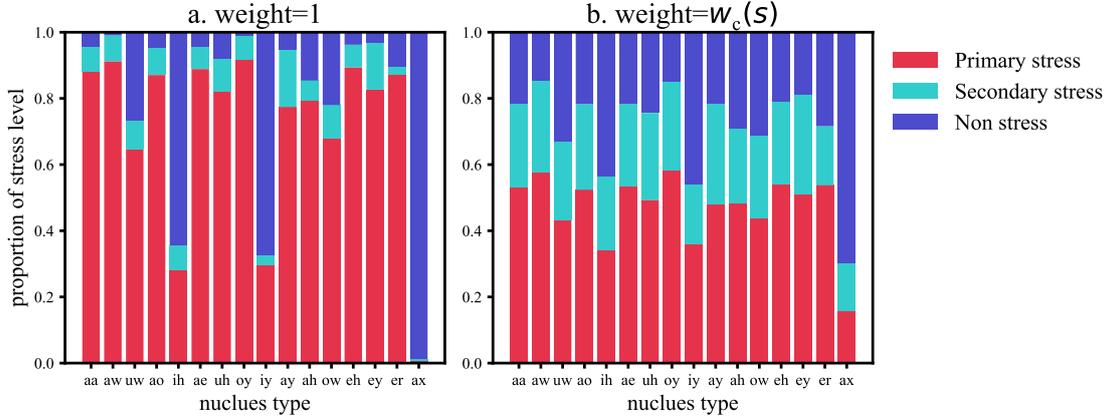

Fig. 3 Proportion of stress level by nucleus type.

## 4.2 Comparison of Models

Models are trained and tested on the Read Data and Generated Data separately. In the following, they are evaluated by the prediction accuracy (i.e., the percentage of syllables whose stress levels are correctly predicted). Table 2 shows the prediction accuracy for the training/testing for various models, features, and dataset settings.

Table 2 Prediction accuracy.

| Data | Model | Syllable Numerical | | Syllable/Nucleus Numerical | | All Features | |
|---|---|---|---|---|---|---|---|
| | | Training | Testing | Training | Testing | Training | Testing |
| Read Data | OR | 68.74% | 68.39% | 68.66% | 68.58% | N.A. | N.A. |
| | RF | 71.42% | 69.35% | 73.94% | 69.64% | N.A. | N.A. |
| | Model-medium | 90.64% | 88.27% | 91.12% | 88.47% | 96.45% (96.78%)* | 94.03% (87.96%) |
| | Model-Large | 93.47% | 88.01% | 93.35% | 88.06% | 98.57% (98.46%) | 94.82% (86.40%) |
| Generated Data | OR | 72.65% | 72.56% | 72.59% | 72.64% | N.A. | N.A. |
| | RF | 78.49% | 76.18% | 78.87% | 76.49% | N.A. | N.A. |
| | Model-medium | 95.19% | 93.64% | 95.61% | 93.53% | 98.01% (97.93%) | 96.84% (93.94%) |
| | Model-Large | 98.06% | 93.42% | 98.17% | 93.65% | 99.37% (99.17%) | 97.89% (94.27%) |

\* Percentage in the parenthesis is the weighted accuracy on stress levels by Equation (2).

From Table 2, the proposed self-attention models outperform conventional classifiers (e.g., OR and RF models). Conventional classifiers ignore the features of syllables around the target syllable, leading to poor performance. In previous studies, when conventional classifiers were applied, some features of syllables proceeding or following the target syllable were included to improve the model accuracy (Tepperman and Narayanan, 2005; Ferrer et al., 2015). The selection of such features was based on empirical evaluations, which caused complexity and inconvenience for data processing and model construction.

Overall, the prediction accuracies for the Generated Data are higher than those for the Read Data. Regarding the self-attention model, increasing the model size improves the performance on the training datasets, while the improvement is limited on the testing datasets. In addition, when numerical syllable features are present ("Syllable Numerical" in Table 2), adding numerical nucleus features ("Syllable/Nucleus Numerical" in Table 2) has a tiny impact on the prediction accuracy, almost for all types of models. This implies that numerical nucleus features add little knowledge to the stress level when the syllable features are used. On the other hand, embedding the nuclei type enables the models to provide higher accuracy ("All features" in Table 2).

For several models proposed in previous studies using various datasets, the best prediction accuracies were around 83% to 94% for double-type classification (i.e., stress and nonstress) and 87%-88% for triple-type classification (i.e., primary stress, secondary stress, and nonstress) (Tepperman and Narayanan, 2005; Ferrer et al., 2015; Yarra et al., 2017; Li et al., 2018; Ramanathi et al., 2019; Yarra and Ghosh, 2022). The models proposed in this study yield 88%-95% and 93%-98% accuracies on the testing sets for the two datasets. It is hard to make a fair comparison because different researchers have employed different datasets, but the high accuracy of our models implies that the self-attention models can be promising for further development.

## 4.3 Investigation of the Results from Self-Attention Models

The Read Data includes a higher variation of speakers and a higher error rate. Thus, analyzing the results from this dataset may provide more insights into the prediction errors the models make. Since there are no remarkable differences in the performances of Model-Medium and Model-Large, we will look into Model-Medium in the following.

Fig. 4 shows the confusion matrix from the Model-Medium for the testing set for Read Data trained with "Syllable Numerical" features. Overall, the non-stress and primary stress objects are mostly correctly labeled, while the secondary stress ones have more errors. Notably, many secondary stresses are labeled as non-stress, probably due to the similarity of pronunciation patterns of the two classes.

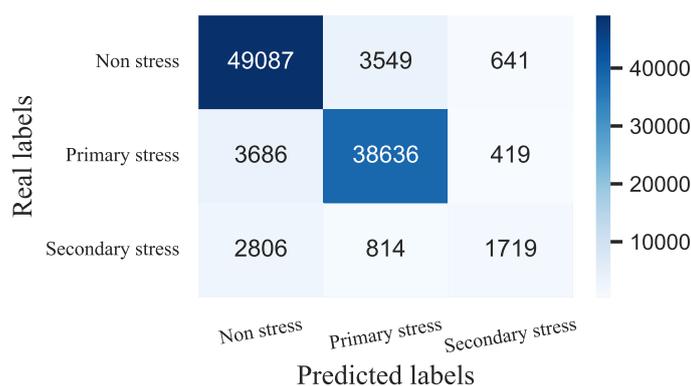

Fig. 4 Confusion matrix from Model-Medium trained with "Syllable Numerical" features.

Fig. 5 shows the confusion matrix for each type of syllable (classified by their nucleus types) trained with "All Features". It can be observed that, although the distribution of stress levels for some types of syllables (e.g., "aw", "oy", "ey") are highly unbalanced, predictions for syllables with rare stress levels are not distorted. This suggests that the weights assigned to syllables have functioned. Similar to the observation from Fig. 4, unstressed syllables are often mislabeled as secondary stress (e.g., "iy", "ih", "ax"). For "ax", some unstressed syllables are mislabeled as primary stress.

|  | iy | | | ih | | | ey | | | eh | | |
|---|---|---|---|---|---|---|---|---|---|---|---|---|
| Non stress | 7123 | 243 | 251 | 11199 | 273 | 995 | 72 | 20 | 22 | 226 | 67 | 49 |
| Primary stress | 151 | 3169 | 19 | 166 | 5183 | 181 | 30 | 2922 | 76 | 144 | 7530 | 212 |
| Secondary stress | 120 | 19 | 231 | 116 | 52 | 1195 | 24 | 25 | 453 | 49 | 51 | 520 |

|  | ae | | | aa | | | ao | | | uh | | |
|---|---|---|---|---|---|---|---|---|---|---|---|---|
| Non stress | 140 | 42 | 28 | 118 | 23 | 16 | 119 | 13 | 18 | 35 | 6 | 2 |
| Primary stress | 132 | 4037 | 101 | 83 | 3200 | 54 | 43 | 2590 | 53 | 14 | 518 | 16 |
| Secondary stress | 15 | 22 | 324 | 22 | 19 | 258 | 14 | 25 | 219 | 5 | 1 | 59 |

|  | ow | | | uw | | | ah | | | ay | | |
|---|---|---|---|---|---|---|---|---|---|---|---|---|
| Non stress | 578 | 40 | 110 | 465 | 26 | 31 | 534 | 42 | 33 | 112 | 8 | 14 |
| Primary stress | 44 | 2093 | 43 | 45 | 1215 | 42 | 88 | 3218 | 36 | 25 | 2065 | 39 |
| Secondary stress | 56 | 17 | 232 | 14 | 18 | 139 | 28 | 12 | 221 | 19 | 33 | 430 |

|  | aw | | | oy | | | ax | | | er | | |
|---|---|---|---|---|---|---|---|---|---|---|---|---|
| Non stress | 6 | 0 | 4 | 1 | 0 | 0 | 29327 | 427 | 333 | 142 | 22 | 4 |
| Primary stress | 4 | 1263 | 7 | 1 | 217 | 6 | 90 | 138 | 0 | 58 | 1376 | 9 |
| Secondary stress | 6 | 9 | 109 | 1 | 3 | 11 | 76 | 0 | 97 | 12 | 4 | 24 |

Real label (y-axis) / Predicted label (x-axis): Non stress, Primary stress, Secondary stress

Fig. 5 Confusion matrix for each type of syllable from Model-Medium on the testing set of the Read Data trained with "All Features".

Nucleus-type embedding is an important step when the self-attention models learn through all features. What the self-attention models learned can be visualized by analyzing the embedded vectors. The embedding vectors for all types of syllables are obtained first. Dimension reduction is applied using Principal Component Analysis and embedded vectors are plotted on the 3-D space (Fig. 6). The plot suggests that the stress patterns of "ow" and "ao", as well as those of "eh" and "er", are similar. The embedding patterns agree with the general pronunciation patterns, indicating that the self-attention models have learned some pronunciation patterns.

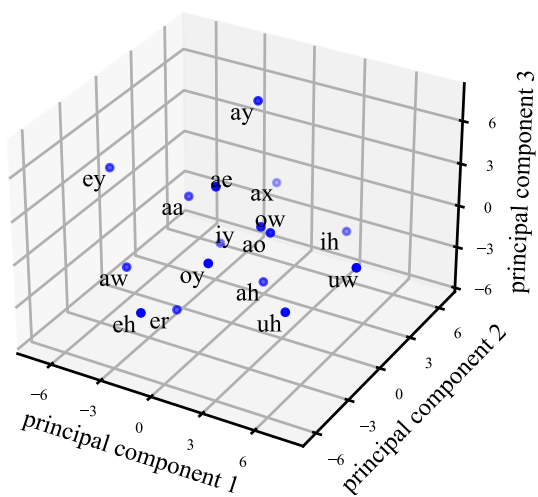

Fig. 6 Visualization of embedded vector for syllable types from the Model-Medium.

## 5 Discussion and Conclusions

In this paper, we presented a self-attention model to identify the stress level of word syllables in English. As presented in Subsection 2.2, this model has several advantages compared with conventional classifiers. First, it in nature takes into consideration all the syllable features around the target syllable in a word. Second, it offers the choice to embed categorical features, such as the syllable or nucleus types. Further analysis suggests that stress patterns seem to have been learned by the model. This model can be used by English learners, or to provide real-time pronunciation assessment for online meetings by incorporating several other modules. Considering the availability of packages for model construction and data processing, this model can be trained and applied in practice efficiently.

Due to the lack of annotated datasets, we adopted a pronunciation dictionary, assuming that the words in the speech audios are correctly stressed. One reading dataset from real readers and two datasets generated using the Azure PA module are applied. The stress levels of the latter datasets are predicted with higher accuracies, probably due to two reasons. First, the Read Data includes speech from 247 readers, and there is a larger variation of speech patterns and more emotional contrasts. These facts made it harder for the models to correctly predict the syllable stress level for all users. Second, the CMU pronunciation dictionary is based on North American pronunciation, and some syllables might be mislabeled because the Read Data includes British and other accents as well. In addition, the datasets used here were collected from speakers with good English pronunciations. This may limit the model application to non-native English learners.

Regarding the feature selection, twelve numerical features over the syllable and the nucleus, as well as the nucleus types, are input for the self-attention models. Our examination shows that, when the features over the syllable are given, the addition of numerical nucleus features has little impact on the model performance. The nucleus type seems to improve the model accuracy, but its impact should be validated with other datasets.

Overall, this study suggests that the self-attention framework can be a promising model for syllable-stress detection. Other features from audio data can be explored in the future. For example, some researchers applied spectral features in their models (e.g., some features extracted from the Fourier transformation). Self-attention models can be conveniently extended to incorporate such features, or even take the spectrum as the input. These topics will be addressed in the future.

# References

Dosovitskiy, A., Beyer, L., Kolesnikov, A., Weissenborn, D., Zhai, X., Unterthiner, T., Dehghani, M., Minderer, M., Heigold, G., Gelly, S., Uszkoreit, J., Houlsby, N., 2020. An Image is Worth 16x16 Words: Transformers for Image Recognition at Scale. Presented at the International Conference on Learning Representations.

Ferrer, L., Bratt, H., Richey, C., Franco, H., Abrash, V., Precoda, K., 2015. Classification of lexical stress using spectral and prosodic features for computer-assisted language learning systems. Speech Communication 69, 31–45. https://doi.org/10.1016/j.specom.2015.02.002

Li, K., Mao, S., Li, X., Wu, Z., Meng, H., 2018. Automatic lexical stress and pitch accent detection for L2 English speech using multi-distribution deep neural networks. Speech Communication 96, 28–36. https://doi.org/10.1016/j.specom.2017.11.003

Ramanathi, M.K., Yarra, C., Ghosh, P.K., 2019. ASR Inspired Syllable Stress Detection for Pronunciation Evaluation Without Using a Supervised Classifier and Syllable Level Features, in: Interspeech 2019. Presented at the Interspeech 2019, ISCA, pp. 924–928. https://doi.org/10.21437/Interspeech.2019-2091

Tepperman, J., Narayanan, S., 2005. Automatic Syllable Stress Detection Using Prosodic Features for Pronunciation Evaluation of Language Learners, in: Proceedings. (ICASSP '05). IEEE International Conference on Acoustics, Speech, and Signal Processing, 2005. Presented at the (ICASSP '05). IEEE International Conference on Acoustics, Speech, and Signal Processing, 2005., IEEE, Philadelphia, Pennsylvania, USA, pp. 937–940. https://doi.org/10.1109/ICASSP.2005.1415269

Vaswani, A., Shazeer, N., Parmar, N., Uszkoreit, J., Jones, L., Gomez, A.N., Kaiser, Ł., Polosukhin, I., 2017. Attention is All you Need, in: Advances in Neural Information Processing Systems. Curran Associates, Inc.

Yarra, C., Deshmukh, O.D., Ghosh, P.K., 2017. Automatic detection of syllable stress using sonority based prominence features for pronunciation evaluation, in: 2017 IEEE International Conference on Acoustics, Speech and Signal Processing (ICASSP). Presented at the 2017 IEEE International Conference on Acoustics, Speech and Signal Processing (ICASSP), IEEE, New Orleans, LA, pp. 5845–5849. https://doi.org/10.1109/ICASSP.2017.7953277

Yarra, C., Ghosh, P.K., 2022. Automatic syllable stress detection under non-parallel label and data condition. Speech Communication 138, 80–87. https://doi.org/10.1016/j.specom.2022.02.001